\documentclass[10pt]{article}

\usepackage{graphicx}
\usepackage{amssymb}
\usepackage{amsmath}
\usepackage{color}
\usepackage{url}
\usepackage{graphicx}        
\usepackage{subfigure}
\usepackage{fancyvrb}
\usepackage{listings}

\usepackage[
    pdftex,
    pdftitle={Unified Residual Evaluation},
    pdfauthor={Matthew G.~Knepley, Jed Brown},
    pdfpagemode={UseOutlines},
    bookmarks,bookmarksopen,bookmarksnumbered={True},
    colorlinks,linkcolor={blue},citecolor={blue},urlcolor={blue}
    ]{hyperref}

\makeatletter
\def\url@leostyle{%
  \@ifundefined{selectfont}{\def\UrlFont{\sf}}{\def\UrlFont{\small\ttfamily}}}
\makeatother
\def\vf{{\bf f}}

\def\class#1{{\tt #1}}
\def\func#1{{\tt #1}}

\newcommand{\blue}{\textcolor{blue}}
\newcommand{\green}{\textcolor{green}}

\usepackage[colorinlistoftodos, textwidth=4cm, shadow]{todonotes}

\renewenvironment{thebibliography}[1]{%
  \begin{oldthebibliography}{#1}%
    \setlength{\itemsep}{0pt} 
}{%
  \end{oldthebibliography}%
}

\usepackage{fullpage}
\usepackage{authblk}
\usepackage{wrapfig}

\begin{document}

\title{Achieving High Performance with Unified Residual Evaluation}
\author[1]{Matthew G. Knepley\thanks{knepley@ci.uchicago.edu}}
\author[2]{Jed Brown\thanks{jedbrown@mcs.anl.gov}}
\author[2]{Karl Rupp\thanks{rupp@mcs.anl.gov}}
\author[2]{Barry F. Smith\thanks{bsmith@mcs.anl.gov}}
\affil[1]{Computation Institute, University of Chicago}
\affil[2]{Mathematics and Computer Science Division, Argonne National Laboratory}
\date{}

\renewcommand\Authands{ and }

\maketitle

\begin{abstract}
We examine residual evaluation, perhaps the most basic operation in numerical simulation. By raising the level of
abstraction in this operation, we can eliminate specialized code, enable optimization, and greatly increase the
extensibility of existing code.
\end{abstract}

\section{Introduction}

With modern partial differential equation (PDE) solver packages,
only performance-critical code the application developers must write is
the ``physics'', that is, the residual form of the PDE they wish to
solve. All the mesh management and algebraic solver code can be contained
in highly optimized libraries. To obtain high performance on modern
computer systems, however, requires minimizing memory motion,
full threading, and fully vectorizable code. Missing even one of the
three could immediately decrease the speed of the code by an order of
magnitude. How can application developers without computer science expertise write
residual code that can be achieve high performance?

At one end, with the
\href{http://www.fenicsproject.org}{FEniCS~project}~\cite{DupHof2003,fenics-web-page,KirbyKnepleyLoggScottTerrel12}
the scientist writes no code at all, just a symbolic representation of
the weak form of the PDE. Internally FEniCS relies on code generation
and separate library routines, rather than a single method for element
traversal and integration to compute the residual. This approach increases the
complexity of the underlying code, however, and has made it too difficult to
support mixed-topology meshes and block-decomposed solvers, as well as to implement implicit material models such as plasticity or real gases or to reuse legacy code.
Other work such as AceGen~\cite{korelc2002multilanguage} has focused on code generation for complex material models across a limited selection of discretizations.
At the other end of the spectrum, with lower-level libraries such as \href{http://libmesh.sourceforge.net}{libMesh}~\cite{libMeshPaper06} and \href{http://dealii.org}{Deal.II}~\cite{bangerth2007deal}, the scientist writes the
required traversal over elements, basis functions, and quadrature
points; hence a new routine is constructed for each application,
and the scientist is required to ensure that it is high performing.
This low-level interface is flexible, allowing the user to implement custom stabilization, choose various implementations of boundary conditions (e.g., slip), implement delta functions, under-integrate certain terms, and countless other forms of messiness in finite element methods.
The user's code is formally independent of the element topology, basis, and quadrature, but the naive representation of basis evaluation is baked into the interface.
\href{https://inlportal.inl.gov/portal/server.pt?open=514&objID=1269&mode=2&featurestory=DA_582160}{MOOSE}~\cite{gaston09,gaston09b,tonks2012object} takes a framework approach, with the goal of creating extensive libraries of physical systems and material models to be used with different classes of physical problems (phase-field, elasticity, fluids, etc.) that are coupled together at runtime into multiphysics simulations.
Most MOOSE users spend their time working on material models and/or coupling various physical systems together, though some create new classes of physical problems.
MOOSE's approach is extremely dynamic, allowing all composition and model configuration to be specified at run-time, but the interface is too fine-grained to vectorize with current compiler technology.
Note that there has been a long tradition in specific problem domains, most notably structural analysis (see FEAP~\cite{taylor2011feap} and commercial packages such as Abaqus) of incorporating extensibility in FEM software via ``elements'' containing the discretization and general structure of the PDE, and ``materials'' providing closures such as the stress-strain relationship.
This organization of material models is compatible with MOOSE (indeed, Abaqus UMAT files can be used directly); but
generic coupling and implementing new models and discretizations is high-effort or unsupported.

We propose an alternative somewhere between these two extremes: the
scientist provides point function evaluations while the library is
responsible for iterating over the mesh elements and quadrature
points. We call this approach ``unified'' because we do not need to
write different iteration schemes in the library for different mesh
entities, dimensions, discretizations, and the like, as is usually done. Our approach offers three benefits.

{\bf Flexibility} Our unified formulation of residual evaluation greatly increases the flexibility of simulation
code. The cell traversal handles arbitrary cell shape (such as simplicial, tensor product, or polygonal) and hybrid
meshes. Moreover, the problem can be posed in any spatial dimension with an arbitrary number of physical fields. It
also accommodates general discretizations, tabulated with a given quadrature rule.

{\bf Extensibility} The library developer needs to maintain only a single method, rather than a code generation
engine or a vast collection of specialized routines, easing language transitions and other environmental
changes. Moreover, extensibility becomes much easier since a new user must master only a small piece of code in order to
contribute to the package. For example, a new discretization scheme could be enabled in a single place in the code.

{\bf Efficiency} A prime motivation for refactoring residual evaluation is to enable stronger, targeted
optimization. In this formulation, only a single routine need be optimized. In addition, the user, application scientist,
or library builder is no longer responsible for proper vectorization, tiling, and other traversal optimization.

We can also enable performance portability of an entire application by recoding a single routine for new hardware. For
example, we have produced OpenCL and CUDA versions of our FEM residual evaluation that achieve excellent performance on
a range of many-core architectures~\cite{KnepleyRuppTerrel13}.

\section{Implementation}

 We will describe in detail the implementation of a generic residual evaluation method for finite element
problems. Our basic operation will be integration of a weak form over a \textit{chunk} of cells. For a CPU, each chunk
is executed in serial and corresponds to classic tiling for cache reuse. On an accelerator, a separate block of threads
will execute each chunk in parallel. Moreover, in our CUDA and OpenCL implementations, a chunk is further subdivided
into \textit{batches} that are processed in serial. The batches are then subdivided into \textit{blocks} that are
processed concurrently by groups of threads. However, the subdivision of the iteration space and resulting traversals,
including those over basis functions and quadrature points, are handled by the library rather than the user, as
discussed below.

At the highest level, we first map all global input vectors to local vectors, using the PETSc \func{DMGlobalToLocal()}
function. The local space contains unknowns constrained by Dirichlet conditions and shared unknowns on process
boundaries that are not present in the global vector used by the solver. In addition, the boundary values are set in the
local solution input vector. For each cell chunk, we extract the input, such as solution FEM coefficients, cell
geometry, and auxiliary field coefficients, then perform the form integration, and insert the result back into the local
residual vector. The local residual then is accumulated into the global residual vector using by 
\func{DMLocalToGlobal()}.

The first serious complication comes when extracting FEM coefficients for a cell chunk. The PETSc unstructured mesh
interface~\cite{KnepleyKarpeev09,AagaardKnepleyWilliams13} uses a Hasse diagram~\cite{HasseDiagram} to represent the
mesh and provides a topological interface independent of the spatial dimension or shape of the constituents, using the
PETSc \class{DMPlex} class. We can ask
for a given breadth-first level in the DAG representing the mesh. For example, cells are leaves of the DAG and thus
have height zero, whereas vertexes have depth zero. These are similar to the concepts of dimension and
codimension, but they arise naturally from the representation. We can prescribe the data layout for any discretization
using a simple size-offset map over all the points in the mesh DAG, called a \class{PetscSection}. For example, a 3D $P_2$
Lagrange element would have one degree of freedom (dof) on each vertex and edge in the mesh. We can then replace continuum geometric
notions with their discrete counterparts to enable generic traversals. In our FEM example, the continuum closure is
replaced by the transitive closure over the mesh DAG, where we stack up dofs as we encounter each point based on the
\class{PetscSection} mapping. Below we show how all these ingredients are used to implement generic coefficient extraction.
\begin{Verbatim}[commandchars={\\\{\}}]
  \green{DM}           dm;
  \green{Vec}          X;
  \green{PetscSection} section;
  \green{PetscScalar} *x, *u;
  \green{PetscInt}     cStart, cEnd, c, n, i, off = 0;

  \blue{DMPlexGetHeightStratum}(dm, 0, &cStart, &cEnd);
  \green{\bf{}for} (c = cStart; c < cEnd; ++c) \{
    \blue{DMPlexVecGetClosure}(dm, section, X, c, &n, &x);
    \green{\bf{}for} (i = 0; i < n; ++i, ++off) u[off+i] = x[i];
    \blue{DMPlexVecRestoreClosure}(dm, section, X, c, &n, &x);
  \}
\end{Verbatim}
A slight complication arises for multiple fields, which the \class{PetscSection} can view as separate maps. Originally,
all dofs associated with a given mesh point are stored contiguously in the global and local vectors. The
\func{DMPlexVecGetClosure()} method reorders the dofs returned so that each field is contiguous.

The code above will not change for different spatial dimensions, number of fields, cell shapes, mesh topologies, or
discretizations. Thus, we can reuse the same code to extract geometric data just by replacing \verb!section! by the
layout of coordinate dofs. After all integration has been carried out, we can use the complementary function
\func{DMPlexVecSetClosure()} to place the residual element vectors into the local residual vector.
\begin{Verbatim}[commandchars={\\\{\}}]
  \green{\bf for} (c = cStart; c < cEnd; ++c) \{
    off = c*cellDof;
    \blue{DMPlexVecSetClosure}(dm, section, F, c, &elemVec[off], ADD_VALUES);
  \}
\end{Verbatim}

We have now reduced the problem of residual evaluation to the integration of a weak form over a small chunk of
cells. We will introduce a simple model of FEM residual evaluation,
\begin{equation}\label{eq:weakForm}
  \mathbf{\phi^T F(u)} \sim \int_\Omega \phi\cdot f_0(u,\nabla u) + \nabla\phi:\vf_1(u,\nabla u) = 0,
\end{equation}
where the pointwise functions $f_0, \vf_1$ capture the problem physics. Discretizing, we have
\begin{equation}\label{eq:weakFormDiscrete}
  \mathsf{F(u)} \sim \sum_e \mathcal{E}^T_e \left[ B^T W f_0(u^q, \nabla u^q) + \sum_k D^T_k W \vf^k_1(u^q, \nabla u^q) \right] = 0,
\end{equation}
where $u^q$ is the vector of field evaluations at the set $q$ of quadrature points on an element, $W$ is the diagonal matrix of quadrature
weights, $B$ and $D$ are basis function matrices that reduce over quadrature points, and $\mathcal{E}_e$ is the element restriction operator.
This approach can be trivially extended to higher orders by adding terms with more pointwise functions. Using this
model, along with automated tabulation of basis functions and derivatives at quadrature points, the user need only
specify physics using pointwise functions similar to the strong form problem. In this way, we decouple the problem
specification from mesh and dof traversal.
The Jacobian of \eqref{eq:weakFormDiscrete} needs only derivatives of the pointwise functions,
\begin{align}\label{eq:weakFormDiscreteJ}
  \mathsf{F'(u)} \sim \sum_e \mathcal{E}_e^T \begin{bmatrix} B^T & \mathbf D^T \end{bmatrix} \mathbf W
  \begin{bmatrix} f_{0,0} & f_{0,1} \\ \vf_{1,0} & \vf_{1,1} \end{bmatrix}
  \begin{bmatrix} B \\ \mathbf D \end{bmatrix} \mathcal E_e, & \quad
      [f_{i,j}] = \begin{bmatrix} \dfrac{\partial f_0}{\partial u} & \dfrac{\partial f_0}{\partial \nabla u} \\[1em]
          \dfrac{\partial \vf_1}{\partial u} & \dfrac{\partial \vf_1}{\partial \nabla u} \end{bmatrix} (u,\nabla u)
\end{align}

For a chunk of elements we may calculate the element residual vectors according to the following pseudo-code.
where we have highlighted opportunities for vectorization:
\vspace{-1.5em}
\begin{tabbing}
\ \ \ \ \=\ \ \ \ \=\ \ \ \ \=\hspace{3in}\=\\
\green{\bf for} chunk in mesh:\\
\>\green{\bf for} c in chunk:             \>\>\>\textit{vectorize over cells}\\
\>\>$u^e$ = \blue{closure}(c, U)\\
\>\>\>$u^q$ = $B u^e$                       \>\textit{vectorize over quadrature points}\\
\>\>\>$\nabla u^q$ = $D u^e$\\
\>\>\>$f_0(u^q, \nabla u^q)$ and $f_1(u^q, \nabla u^q)$\\
\>\>$f^e = B^T W f_0 + D^T W f_1$             \>\>\textit{vectorize over basis functions}\\
\>\>\blue{closure}(c, F) += $f^e$
\end{tabbing}

\begin{wrapfigure}[15]{r}{0.6\textwidth}
  \vspace{-1em}
  \centering
  \includegraphics[width=0.5\textwidth]{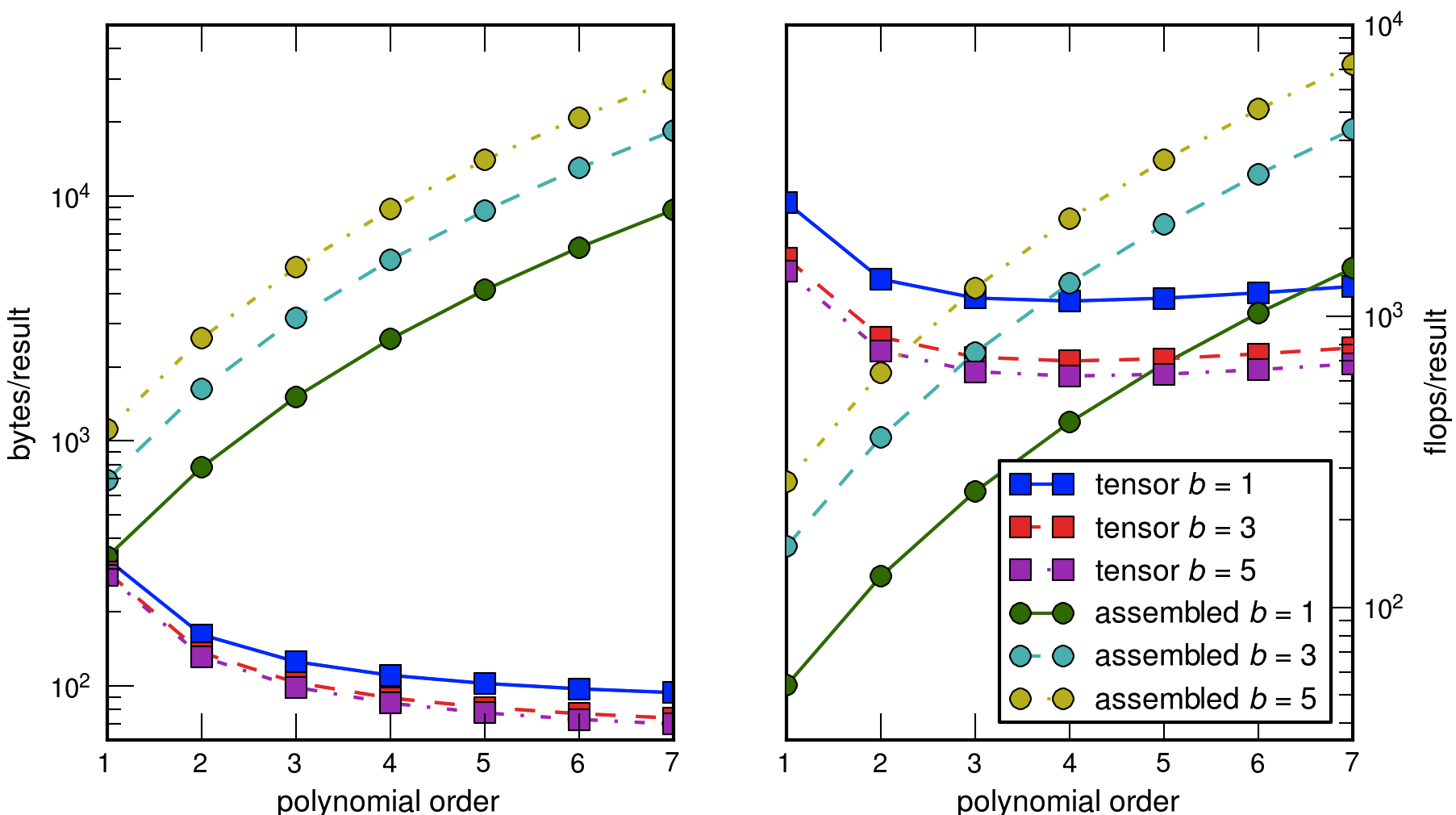}
  \caption{Memory bandwidth and flops per dof to apply a Jacobian from $Q_k$ discretization of a $b$-variable PDE system using an assembled matrix versus matrix-free exploiting the tensor product structure in \eqref{eq:weakFormDiscreteJ}.} 
  \label{fig:assembly}
\end{wrapfigure}

Note that this is the only code that must be ported to new architectures in order to realize all the gains cited
above. In recent work~\cite{KnepleyRuppTerrel13}, the OpenCL implementation of cell chunk integration was used to do
residual evaluation on several accelerator architectures, including Nvidia, ATI, and Intel MIC. We employed mixed
vectorization over both quadrature points and basis functions. For the Nvidia GTX580 in particular, we achieve almost
300 GF/s for a first-order discretization. In Fig.~\ref{fig:assembly} (see~\cite{brown2010ens}) we see that
unassembled operator application greatly reduces memory bandwidth requirements while being competitive in terms of flops for all but lowest order discretizations.
Preconditioning will also be required, for which low-order embedded methods are an unintrusive approach.
Experiments with the dual order implementation \href{https://github.com/jedbrown/dohp}{Dohp} have shown that a generic variable-order tensor-product implementation can be at least as efficient at all orders as the libMesh and Deal.II implementations, which are based on traditionally-assembled element routines.

We are extending this model to pointwise Riemann solvers for hyperbolic conservation
laws. The integration would now take place over faces instead of cells, which means we must allow an input cell height
for our traversal. Moreover, we update the \textit{support} of the cell, which is the dual of its closure. The traversal
also becomes more complicated when using reconstruction, since this requires a cell traversal as well. However, it
seems clear that the model can be generalized to accommodate these changes while maintaining both its simplicity and
its efficiency.

\vskip5em

{\bf Acknowledgments} MGK acknowledges partial support from DOE Contract DE-AC02-06CH11357 and NSF Grant OCI-1147680.
JB and BFS were support by the U.S. Department of Energy, Office of Science, Advanced Scientific Computing Research, under Contract 
DE-AC02-06CH11357.

\bibliographystyle{plain}
{\small 
\bibliography{paper,petsc,petscapp}
}

{\bf Government License.} The submitted manuscript has been created by UChicago Argonne, LLC,
Operator of Argonne National Laboratory (``Argonne'').
Argonne, a U.S. Department of Energy Office of Science laboratory, is
operated under Contract No. DE-AC02-06CH11357. The U.S. Government
retains for itself, and others acting on its behalf, a paid-up
nonexclusive, irrevocable worldwide license in said article to reproduce,
prepare derivative works, distribute copies to the public, and perform
publicly and display publicly, by or on behalf of the Government.

\end{document}